\documentclass[sn-mathphys,pdflatex]{sn-jnl}

\jyear{2021}%

\theoremstyle{thmstyleone}%
\theoremstyle{thmstyletwo}%

\theoremstyle{thmstylethree}%

\begin{document}

\title[Deviation of geodesics and particle trajectories in a gravitational wave of the Bianchi type VI universe]{Deviation of geodesics and particle trajectories in a gravitational wave of the Bianchi type VI universe}




\author*[1,2]{\fnm{Konstantin} \sur{Osetrin}}\email{osetrin@tspu.edu.ru}

\author[1]{\fnm{Evgeny} \sur{Osetrin}}\email{evgeny.osetrin@tspu.edu.ru}
\equalcont{These authors contributed equally to this work.}

\author[1]{\fnm{Elena} \sur{Osetrina}}\email{elena.osetrina@tspu.edu.ru}
\equalcont{These authors contributed equally to this work.}

\affil*[1]{\orgdiv{Center for Mathematical and Computer Physics}, \orgname{Tomsk State Pedagogical University}, \orgaddress{\street{Kievskaya str. 60}, \city{Tomsk}, \postcode{634061}, 
\country{Russia}}}

\affil[2]{
\orgname{National Research Tomsk State University}, \orgaddress{\street{Lenina pr. 36}, \city{Tomsk}, \postcode{634050}, 
\country{Russia}}}



\abstract{
For the Bianchi type VI universe, exact solutions of the equation of geodesic deviation in a strong primordial gravitational wave in a privileged coordinate system are obtained.
The solutions refer to Shapovalov's gravitational-wave models of spacetime and allow the existence of complete integrals of the Hamilton-Jacobi equation for test particles. For all the solutions obtained, the analytical form of the tidal acceleration vector in a strong primordial gravitational wave is obtained. An explicit form of the coordinate transformation, an explicit form of the metric of the primordial gravitational wave of the Bianchi type VI universe, and the form of the tidal acceleration vector in the laboratory synchronous coordinate system are obtained. The synchronous coordinate system is associated with a freely falling observer and allows the observer to separate time and spatial coordinates, as well as to synchronize time at different points in space. The presented mathematical approach can be applied both in the general theory of relativity and in modified theories of gravity.
}





\maketitle

\section{Introduction}\label{sec0}

The development of gravitational-wave astronomy \cite{PhysRevLett.116.061102,PhysRevX.9.031040,PhysRevX.11.021053}, as a new way of obtaining astrophysical information, requires the development of both mathematical methods and obtaining new exact mathematical models in this area. At present, the main calculation methods in the field of analysis of gravitational-wave signals are approximate and numerical methods. So, for the analysis and recognition of gravitational-wave signals from the merger of black holes and neutron stars, bases of "pattern" of numerical models of gravitational waves of a similar type for their direct detection have been formed and are being developed. On the other hand, the observation of long-range gravitational waves in direct observation is difficult, but possible due to the secondary effects of the impact of these waves on astrophysical objects, for example, perturbation of the periods of pulsars or perturbation of the gravitational lensing of distant galaxies, but exact mathematical models are needed to calculate such effects. For example, primordial gravitational waves could make a specific contribution to the electromagnetic microwave background of the universe \cite{Bennett2013}, acting on the primary plasma due to tidal accelerations, and also lead to the emergence of secondary gravitational waves due to amplification of density fluctuations, which in turn could give resonances in the observed stochastic gravitational wave background. The calculation of such models is quite complicated, but, nevertheless, 
it is possible to construct exact models of primordial gravitational waves with finding the exact trajectories of test particles, obtaining exact solutions to the geodesic deviation equations and finding the explicit form of tidal accelerations, which allows further calculation of the effect of a gravitational wave on astrophysical objects, including primary plasma, dust or radiation and calculate the additional electromagnetic or gravitational radiation generated as a result \cite{Domenech2021398}. In addition, dynamic density fluctuations caused by primordial gravitational waves can lead to the formation of primordial black holes \cite{Saito200916,Saito2010867}, which is an important astrophysical effect. The generation of black holes by primordial gravitational waves also requires the creation of appropriate mathematical models to describe such phenomena.

When considering the early stages of the dynamics of the universe, the methods of perturbation theory in a homogeneous and isotropic universe Friedmann-Lemaître-Robertson-Walker 
are often used \cite{Ma19957,MUKHANOV1992203}.
On the other hand, obtaining exact models of primordial gravitational waves is also possible in Bianchi's nonisotropic models of universes, which makes it possible to determine their influence on astrophysical objects and study astrophysical models of gravitational waves in the presence of dust, radiation, and plasma, which serves as a versatile verification of existing and new mathematical approaches to describing the early history of the universe.
To clarify the scenarios for the development of the universe at the initial stages, when the universe could be nonisotropic, it is useful to have exact models of primordial gravitational waves for various nonisotropic Bianchi models, which could make a different contribution both to the currently observed cosmic microwave background and to the gravitational-wave stochastic background of the universe, which one could try to verify from observational data (see 
\cite{Domenech2021398}).

In this paper, we complete the classification of exact models of primordial gravitational waves based on type III Shapovalov wave spacetimes for Bianchi universes. Recall that, according to the previous analysis \cite{OsetrinHomog312002,OsetrinHomog2006}), type III Shapovalov gravitational-wave spacetimes admit Bianchi universes of types IV, VI, and VII.
Previously, we have already obtained exact models and found exact solutions for test particle trajectories, exact solutions for geodesic deviation equations, and calculated tidal accelerations in primordial gravity wave models for Bianchi type IV universes \cite{Osetrin2022EPJP856} and Bianchi type VII universes \cite{https://doi.org/10.48550/arxiv.2206.15234}. 

In this paper, we consider exact models of primordial gravitational waves in Bianchi type VI universes. Consideration of gravitational wave models based on Shapovalov spacetimes is based on the observational fact that the propagation velocity of gravitational waves is equal to the speed of light \cite{Abbott2017PRL161101}, which allows using privileged coordinate systems with the selection of wave variables along which the spacetime interval vanishes and where an exact integration of the equations of test particles in the Hamilton-Jacobi formalism is possible. 

An explicit form of transformations from 
privileged coordinate systems (where the wave variable is used and it is possible to exactly integrate the equations of motion of test particles in the Hamilton-Jacobi formalism) 
into synchronous reference systems (where the time variable and spatial variables are separated and time synchronization is possible at different points in space) is constructed. An explicit form of the gravitational wave metric of Bianchi type VI universes in a synchronous frame of reference associated with a freely falling observer is obtained. 
Exact solutions for the trajectories of test particles and exact solutions of geodesic deviation equations are obtained, tidal accelerations in a gravitational wave are found both in a privileged coordinate system and in a synchronous reference system.

\section{Stäckel  spaces and deviation of geodesics}

For completeness, we recall a number of statements from the Hamilton-Jacobi formalism, the theory of Stäckel spaces, Shapovalov spaces, and other information necessary to understand the content.

The equation 
of test particles in a gravitational field in the Hamilton-Jacobi formalism has the form (see \cite{LandauEng1}):
\begin{equation} 
g^{{\alpha}{\beta}}\frac{\partial S}{\partial x^{\alpha}}\frac{\partial S}{\partial x^{\beta}}=m^2c^2
,
\qquad
{\alpha},{\beta},{\gamma},{\delta}=0,...(n-1),
\label{HJE}
\end{equation} 
where $m$ is the test particle mass, $c$ is the speed of light, $g^{{\alpha}{\beta}}$ is the space metric, $S$ is the test particle action function, $n$ is the space dimension. In what follows, we will set the speed of light $c$ equal to unity.

The "complete integral" of the Hamilton-Jacobi equation (\ref{HJE}) is the solution 
of this equation $S(x^{\alpha},\lambda_{\beta})$, which contains $n$ independent constants 
$\lambda_{\beta}$,  
i.e. such that the following condition is satisfied:
\begin{equation}
\left\|
\frac{\partial{}^2 S}{\partial x^{\alpha} \partial \lambda_{\beta}}
\right\|
\ne 0
.
\end{equation}

If the complete integral of the equation (\ref{HJE})  is found, then the trajectory of the test particle can be found from equations of the form:
\begin{equation}
\label{MovEqu}
\frac{\partial S (x^{\alpha},\lambda_{\beta})}{\partial \lambda_{\gamma}}=\sigma_{\gamma},
\qquad
\tau=S (x^{\alpha},\lambda_\beta)\Bigl\|_{m=1},
\end{equation}
where $\lambda_{\gamma}$, $\sigma_{\gamma}$ are independent constant parameters determined by the initial data of the test particle motion, $\tau$ is the proper time of the particle.

The presence of the complete integral of the equation (\ref{HJE}) also makes it possible to find coordinate transformations for the transition to synchronous reference systems \cite{LandauEng1}, where time and spatial coordinates are separated from each other (which is important for physical calculations) and time synchronization at different points of space is possible.

One of the main methods for solving the Hamilton-Jacobi equation (\ref{HJE}) is the method of separation of variables in privileged coordinate systems. For the first time this issue began to be studied by Paul Stäckel (see \cite{Stackel1897145}), in honor of which the spaces that allow complete separation of variables in the Hamilton-Jacobi equation (\ref{HJE}) were called "Stäckel  spaces". The theory of Stäckel spaces was built by the efforts of many researchers and received its final form in the works of 
\mbox{Vladimir} \mbox{Shapovalov} \cite{Shapovalov1978I,Shapovalov1978II,Shapovalov1979}), 
 where a classification of such spaces was carried out and a metric structure was obtained in privileged coordinate systems (where separation of variables is allowed) for all types of "Stäckel  spaces". In the works of Shapovalov, for the first time, types of spaces were distinguished that allowed the separation of variables with the separation of isotropic variables, along which the interval vanished, we call such models of spacetime ''Shapovalov wave spaces'' \cite{Osetrin2020Symmetry}. As gravitational wave observations confirm, gravitational waves propagate at the speed of light \cite{Abbott2017PRL161101} and, thus, the selection of wave variables along which the spacetime interval vanishes has physical grounds in the study of gravitational waves. Thus, from the point of view of physics, Shapovalov spaces are related to wave models of spacetime, i.e. describe gravitational waves, which is of particular importance for the development of mathematical methods of gravitational-wave astronomy.
 
Stäckel spaces are classified according to the type of so-called "complete sets" of commuting Killing vectors and Killing tensors of the second rank that they allow, which determine sets of integrals of motion of a test particle, linear and quadratic in momenta.

The possibility of constructing a solution of the Hamilton-Jacobi equations in the form of a complete integral also makes it possible to obtain an exact solution for the geodesic deviation equations:
\begin{equation}
\label{deviation}
\frac{D^2 \eta^{\alpha}}{{d\tau}^2}=
 R^{\alpha}{}_{{\beta}{\gamma}{\delta}}u^{\beta} u^{\gamma}\eta^{\delta} .
\end{equation}
Here
$u^{\alpha}$ is the four-velocity of the test particle on the base geodesic line,
$D$ is the covariant derivative,
$\eta^{\alpha}$ is the geodesic deviation vector, $R^{\alpha}{}_{{\beta}{\gamma}{\delta}}$ is the Riemann curvature tensor.

The deviation of geodesics underlies the physical content of the metric theories of gravity, since it sets the mutual motion of neighboring test particles in a gravitational field, which can actually be observed. Direct detection of the gravitational field and gravitational waves is based on the observation of geodesic deviations. Therefore, obtaining exactly solvable models in this area is of undoubted mathematical and physical interest.

For the 4-velocity of a test particle on a base geodesic satisfying the condition
\begin{equation}
\label{Norm}
g^{{\alpha}{\beta}}u_{\alpha}u_{\beta}=1
,
\end{equation}
in the Hamilton-Jacobi formalism we have the following representation
\begin{equation}
u_{\alpha}=\frac{\partial S}{\partial x^{\alpha}}\biggr\|_{m=1}
.
\end{equation}
Then the 4-velocity components take the following form
\begin{equation}
u_{\alpha}=u_{\alpha}(x^{\gamma},\lambda_1,...,\lambda_{(n-1)})
.
\end{equation}
The proper time $\tau$ of the particle on the base geodesic takes the form
\begin{equation}
\tau =S(x^{\gamma},\lambda_1,...,\lambda_{(n-1)})\Bigr\|_{m=1}
.\end{equation}

It was previously shown by variational methods \cite{Bazanski19891018} that the presence of a complete integral of
the Hamilton-Jacobi equation (\ref{HJE}) allows one to obtain the solution
geodesic deviation equations (\ref{deviation}) as a solution to the system of equations of the following form
\begin{equation}
\label{Deviation1}
\eta^{\gamma}\,
\frac{\partial u_{\gamma}(x^{\alpha},\lambda_{k})}{\partial\lambda_{i}}
+\rho_{j}\frac{\partial^2 S(x^{\alpha},\lambda_{k})}{\partial\lambda_{i}\partial\lambda_{j}}=\vartheta_{i}
,
\end{equation}
\begin{equation}
\label{Deviation2}
u_{\gamma}(x^{\alpha},\lambda_{i})\,\eta^{\gamma}=0
,
\quad
{i},{j},{k} = 1\ldots 3;
\quad
{\alpha},{\beta},{\gamma}=0\ldots 3,
\end{equation}
where $\lambda_{i}$, $\rho_{i}$, $\vartheta_{i}$ are independent constant parameters.

The parameters $\lambda_{i}$ set the initial or boundary values of the test particle velocities on the base geodesic, and the constants $\rho_{i}$, $\vartheta_{i}$ set the initial or boundary values of the relative position and velocity of the particle on the neighboring geodesic with the deviation vector $\eta^{\alpha}$.

\section{Bianchi-Shapovalov wave universes}

Shapovalov wave spacetimes that allow separation of wave variables in privileged coordinate systems are classified according to the number of commuting Killing vectors they allow in ''complete set''. For spacetimes there are three types of Shapovalov spaces: Type I (one Killing vector), Type II (two commuting Killing vectors) and III type (three commuting Killing vectors).

When studying cosmological models of the early stages of the universe in order to construct mathematical models of primordial gravitational waves, it becomes necessary to study Bianchi universes, which are spatially homogeneous, but non-isotropic models of spacetime and allow exact gravitational-wave solutions. The Bianchi universes admit the existence of three-parameter groups of motion of spacetime with space-like orbits. 

Thus, in order to construct exact mathematical models of primordial gravitational waves in Shapovalov spacetimes, we come to the problem of identifying models that admit symmetries of Bianchi universes.
It was shown earlier that type III Shapovalov wave spacetimes allow Bianchi type IV, VI and VII 
 universes \cite{OsetrinHomog312002,OsetrinHomog2006}, and type II Shapovalov spaces allow type III Bianchi universes \cite{OsetrinHomog212020}. For Shapovalov spaces of type I, this classification has not yet been carried out.
Thus, we currently have four types of Bianchi universes for which we can construct exact models of primordial gravitational waves in Shapovalov spacetimes, that allow exact integration of the test particles equations.

In this paper, we complete the classification of exact solutions of geodesic deviation equations in type III Shapovalov spacetimes for a gravitational wave in the Bianchi universe of type VI
whose metric in the privileged coordinate system
can be represented in the following form \cite{OsetrinHomog2006}:
\begin{equation}
{ds}^2=2dx^0dx^1
-
\frac{1}{{\sin^2\!{\alpha}}}
\left[
{\left(x^0\right)}^{2{p}}{(dx^2)}^2
+2\,{\cos{({\alpha})}} \,{\left(x^0\right)}^{{p}+{q}}{dx^2}{dx^3}
+{\left(x^0\right)}^{2{q}}{(dx^3)}^2
\right]
\label{Metric}
,
\end{equation}
where $x^0$ is the null (wave) variable,
constants ${p}$, ${q}$ and ${\alpha}$ are independent parameters of the model and
\begin{equation}
0<{\alpha}<\pi
.\end{equation}
The spacetimes (\ref{Metric}) is plane-wave since it admits a covariantly constant vector~$K$:
\begin{equation}
\nabla_{\beta} K_{\alpha}=0
\quad
\to
\quad
K_{\alpha}=\bigl( K_0,0,0,0 \bigl)
,\qquad
 K_0=\mbox{const}
.
\end{equation}
The spacetimes  (\ref{Metric}) admits a group of motions of spatial homogeneity with Killing vectors $X_{(1)}$, $X_{(2)}$, and $X_{(3)}$:
\begin{equation}
X^{\alpha}_{(1)}=\bigl(0,0,1,0\bigr),
\quad
X^{\alpha}_{(2)}=\bigl(0,0,0,1\bigr),
\quad
X^{\alpha}_{(3)}=\bigl(-x^0,\, x^1,\, {p} x^2,\, {q} x^3\bigr)
.\end{equation}
An additional fourth Killing vector
has the form
\begin{equation}
X^{\alpha}_{(0)}=\bigl(0,1,0,0\bigr)
,
\qquad
g_{{\alpha}{\beta}}X^{\alpha}_{(0)}X^{\beta}_{(0)}=0
.
\end{equation}
The commutation relations for the Killing vectors have the following form:
\begin{equation}
\left[X_{(0)},X_{(1)}\right]=0
,\qquad
\left[X_{(0)},X_{(2)}\right]=0
,\qquad
\left[X_{(0)},X_{(3)}\right]=X_{(0)}
,\qquad
\end{equation}
\begin{equation}
\left[X_{(1)},X_{(2)}\right]=0
,\qquad
\left[X_{(1)},X_{(3)}\right]={p} X_{(1)}
,\qquad
\left[X_{(2)},X_{(3)}\right]={q} X_{(2)}
.
\end{equation}
The Killing vectors 
$X_{(1)}$, $X_{(2)}$ and $X_{ (3)}$ generate a 3-dimensional subgroup of the spatial homogeneity of the type VI Bianchi universe.

The Riemann curvature tensor for the type VI plane-wave Bianchi cosmological model with the metric (\ref{Metric}) in the privileged coordinate system has the following nonzero components:

\begin{equation}
{ R }_{0202} = \frac{
\sigma^2\left[
4{p}(1-{p})-({p}-{q})^2
\right]
+({p}-{q})^2
}{4 {\sigma}^4}
\,{\left(x^0\right)}^{2 {p} -2}
\end{equation}
\begin{equation}
{ R }_{0302} = \frac{{a}
\left[
\sigma^2({p}+{q})({p}+{q}-2)-({p}-{q})^2
\right]
}{4 {\sigma}^4}
\,{\left(x^0\right)}^{{p} +{q} -2}
\end{equation}
\begin{equation}
{ R }_{0303} = \frac{
\sigma^2\left[
4{q}(1-{q})-({p}-{q})^2
\right]
+({p}-{q})^2
}{4 {\sigma}^4}
\,
{\left(x^0\right)}^{2 {q} -2}
\end{equation}
The Ricci tensor ${ R }_{{\alpha}{\beta}}={ R }^{\gamma}{}_{{\alpha}k{\beta}}$ in the privileged coordinate system has one non-zero component:

\begin{equation}
{ R }_{00} = -\frac{
\sigma^2({p}+{q})({p}+{q}-2)
+({p}-{q})^2
}{2 {\sigma}^2 {\left(x^0\right)}^2}
.
\end{equation}
The scalar curvature ${ R }$ vanishes.

Weyl conformal curvature tensor
 in the privileged coordinate system has the following non-zero components

\begin{equation}
{\rm C}_{0202} = -\frac{({p} -{q} ) \left[
{\sigma}^2(2{p}-1)-{p}+{q}
\right]}{2 {\sigma}^4}
\,
{\left(x^0\right)}^{2 {p} -2}
,\end{equation}
\begin{equation}
{\rm C}_{0302} = -\frac{{a} ({p} -{q} )^2 }{2 {\sigma}^4}
\,
{\left(x^0\right)}^{{p} +{q} -2}
,\end{equation}
\begin{equation}
{\rm C}_{0303} = \frac{({p} -{q} )
\left[
{\sigma}^2(2{q}-1)+{p}-{q}
\right]
}{2 {\sigma}^4}
\,
 {\left(x^0\right)}^{2 {q} -2}
.\end{equation}
Thus, for 
${p}={q}$ the model under consideration degenerates and the spacetime becomes conformally-flat.

\section{Gravitational wave for Bianchi type VI models in Einstein's theory of gravitation}

In this paper, we will consider gravitational wave models for the Einstein equations in vacuum.
Note that the mathematical approach considered in the paper allows obtaining exact models for various types of matter:
radiation or dust 
\cite{OsetrinDust2016,OsetrinRadiation2017},
electromagnetic field 
\cite{Obukhov202284,Bagrov19881141},
scalar fields 
\cite{OsetrinScalar2018,Obukhov2022632,Obukhov2022142,Obukhov2021134,Obukhov2021183}, 
modified theories of gravity 
\cite{Odintsov2007,Odintsov2011,Capozziello2011,Odintsov2017}),
etc.

Einstein's equations with cosmological constant $\Lambda$ in vacuum

\begin{equation}
R_{{\alpha}{\beta}}=\Lambda g_{{\alpha}{\beta}},
\end{equation}
for the metric (\ref{Metric}) give the following conditions only:

\begin{equation}
\label{GeneralSigma}
\Lambda=0
,\qquad
{\sin^2\!{\alpha}}=
\frac{({p} -{q} )^2}{1-({p} +{q} -1)^2}
,\end{equation}
or, equivalently, for the constant ${\alpha}$ we get

\begin{equation}
\label{GeneraAlpha}
{\cos^2\!{\alpha}}=
2
\,
\frac{
 {p} ({p} -1)+ {q} ({q} -1)
}{({p} +{q} -2) ({p} +{q} )}
.
\end{equation}

Of the three 
parameters of the gravitational wave model, only two independent parameters  remain: ${p}$ and ${q}$.
The relations for the angular parameter~${\alpha}$ due to 
(\ref{GeneralSigma}) and (\ref{GeneraAlpha}), lead to conditions for the parameters of the model, determining the allowable range for the parameters ${p}$ and ${q} $. Due to the requirement that the metric be non-degenerate
and from the Einstein equations we obtain the condition ${p}\ne{q}$, i.e. the model under consideration cannot lead to a conformally flat space-time.

Due to the symmetry of the permutation of the variables $x^2$ and $x^3$ and taking into account the condition ${p}\ne{q}$, we can put ${p}>{q}$ for definiteness without loss of generality, then we can distinguish three admissible ranges of parameters
${p}$ and ${q}$:

\begin{equation}
0<{p} \leq 1
,\qquad
\frac{1-\sqrt{1+4 {p} (1- {p})}}{2}
\leq {q} <{p}
\label{ParameteRange1}
,
\end{equation}
\begin{equation}
 1<{p}<
\frac{1+\sqrt{2}}{2}
,\quad
\frac{1-\sqrt{1+4 {p} (1- {p})}}{2}
\leq {q} \leq
\frac{1+\sqrt{1+4 {p} (1- {p})}}{2}
\label{ParameteRange2}
   ,
\end{equation}
\begin{equation}
{p} =\frac{1+\sqrt{2}}{2}
,\qquad
{q}=\frac{1}{2}
\label{ParameteRange3}
.\end{equation}
Moreover, for the indicated intervals, the following relation holds:
\begin{equation}
0<{p}+{q}<2
.\end{equation}
For all admissible ranges of parameters (\ref{ParameteRange1})-(\ref{ParameteRange3}) the condition ${p}>0$ is satisfied, while the parameter ${q}$ can take both positive and negative values.

The solution of Einstein's vacuum equations for the considered gravitational wave can be represented in another parametric form, using explicitly only two independent angular parameters ${\alpha}$ and ${{\beta}}$, as follows:

\begin{equation}
{ds}^2=2dx^0dx^1
-
\frac{1}{{\sin^2\!{\alpha}}}
\left[
{\left(x^0\right)}^{2{p}}{(dx^2)}^2
+2\,{\cos{({\alpha})}}\, {\left(x^0\right)}^{{p}+{q}}{dx^2}{dx^3}
+{\left(x^0\right)}^{2{q}}{(dx^3)}^2
\right]
\label{Metric3}
,
\end{equation}
\begin{equation}
{{p}}=\frac{1}{2}\,
\left(
1+\cos{\beta} + \sin{\alpha} \sin{\beta}
\right)
\label{Metric3B}
,\end{equation}
\begin{equation}
{{q}}=\frac{1}{2}\,
\left(
1+\cos{\beta} - \sin{\alpha} \sin{\beta}
\right)
\label{Metric3C}
,\end{equation}
\begin{equation}
0<{\alpha} <\pi
,\qquad
0<{{\beta}}<\pi
\label{Metric3D}
.\end{equation}
Note that the Riemann curvature tensor, taking into account the Einstein equations, can vanish identically (i.e., the considered spacetime becomes flat) only when
${\alpha}={{\beta}}=\pi/2$, and in this case 
${{p}}={{q}}=1$.

\section{Integration of the Hamilton-Jacobi equation for test particles (
case 
$p+q\ne 1$
\&
 $p,q\ne 1/2$)}

The test particle Hamilton-Jacobi equation for the metric (\ref{Metric3}) in the privileged coordinate system under consideration allows integration by the separation of variables method, when the action function $S$ for the test particle can be written in the privileged coordinate system in the ''separated'' form

\begin{equation} 
S={\phi_0}\left(x^0\right) +{\lambda_1} x^1+{\lambda_2} x^2+{\lambda_3}x^3
,\end{equation} 
where $\lambda_{i}$ are the constant parameters determined by the initial conditions.

The Hamilton-Jacobi equation (\ref{HJE}) takes the form (the test particle mass is set to unity, $m=1$):
\begin{equation} 
 0 = -2\, {\cos{({\alpha})}} \, {\lambda_2} {\lambda_3}\, {\left(x^0\right)}^{-{p} -{q} }-2 {\lambda_1} {\phi_0}'\left(x^0\right)-{\lambda_2}^2 {\left(x^0\right)}^{-2 {p} }-{\lambda_3}^2 {\left(x^0\right)}^{-2 {q} }+1 
, \end{equation} 
hence for the function $ \phi_0\left(x^0\right)$ we obtain the equation

$$
 \phi_0{}' = -\frac{2 {\cos{({\alpha})}} {\lambda_2} {\lambda_3} {\left(x^0\right)}^{-{p} -{q} }+{\lambda_2}^2 {\left(x^0\right)}^{-2 {p} }+{\lambda_3}^2 {\left(x^0\right)}^{-2 {q} }-1}{2 {\lambda_1}} 
, $$
and, apart from special cases (when ${p}$ or ${q}$ equals $1/2$ or ${p}+{q}$ equals $1$), we get
$$
 \phi_0 
 = 
 -\frac{{x^0} 
  }{2 {\lambda_1}} 
 \left(-\frac{2\, {\cos{({\alpha})}} \, {\lambda_2} {\lambda_3} {\left(x^0\right)}^{-{p} -{q} }}{{p} +{q} -1}+\frac{{\lambda_2}^2 {\left(x^0\right)}^{-2 {p} }}{1-2 {p} }+\frac{{\lambda_3}^2 {\left(x^0\right)}^{-2 {q} }}{1-2 {q} }-1\right)
. $$
Special cases of solutions of the Hamilton-Jacobi equation, when ${p}$ or ${q}$ are equal to $1/2$, or $({p}+{q})=1$, will be considered separately below.

The equations for the trajectories of motion of test particles (\ref{MovEqu}) can be written 
in the following form

$$
 \tau = 
 -\frac{
 {x^0} 
 }{2 {\lambda_1}}
 \left(-\frac{2 \,{\cos{({\alpha})}} \,{\lambda_2} {\lambda_3} {\left(x^0\right)}^{-{p} -{q} }}{{p} +{q} -1}+\frac{{\lambda_2}^2 {\left(x^0\right)}^{-2 {p} }}{1-2 {p} }+\frac{{\lambda_3}^2 {\left(x^0\right)}^{-2 {q} }}{1-2 {q} }-1\right)
$$
$$
 +{\lambda_1} {x^1}+{\lambda_2} {x^2}+{\lambda_3} {x^3} 
 +{\sigma_0}
, $$
$$
 0 = 
 \frac{{x^0} 
  }{2 {\lambda_1}^2}
 \left(-\frac{2\, {\cos{({\alpha})}} \,{\lambda_2} {\lambda_3} {\left(x^0\right)}^{-{p} -{q} }}{{p} +{q} -1}+\frac{{\lambda_2}^2 {\left(x^0\right)}^{-2 {p} }}{1-2 {p} }+\frac{{\lambda_3}^2 {\left(x^0\right)}^{-2 {q} }}{1-2 {q} }-1\right)
 -{\sigma_1}+{x^1} 
, $$
$$
 0 = \frac{{\cos{({\alpha})}}\, {\lambda_3} {\left(x^0\right)}^{-{p} -{q} +1}}{{\lambda_1} ({p} +{q} -1)}-\frac{{\lambda_2} {\left(x^0\right)}^{1-2 {p} }}{{\lambda_1}-2 {\lambda_1} {p} }-{\sigma_2}+{x^2} 
, $$
$$
 0 = \frac{{\cos{({\alpha})}}\, {\lambda_2} {\left(x^0\right)}^{-{p} -{q} +1}}{{\lambda_1} ({p} +{q} -1)}-\frac{{\lambda_3} {\left(x^0\right)}^{1-2 {q} }}{{\lambda_1}-2 {\lambda_1} {q} }-{\sigma_3}+{x^3} 
, $$
where the constants $\sigma_{i}$ can be set equal to zero by choosing the origin of the coordinate variables and the proper time of the particle $\tau$.

Solving the test particle trajectory equations in the privileged coordinate system then gives
($p,q\ne 1/2$ and $(p+q)\ne 1$):
\begin{equation} 
\label{GeneralX0}
x^0 ( {\tau} ) = {\lambda_1} {\tau} 
, \end{equation} 
\begin{equation} 
\label{GeneralX1}
x^1 ( {\tau} )  = 
\frac{\tau}{2 {\lambda_1}} 
+\frac{ {\cos{({\alpha})}}\, {\lambda_2} {\lambda_3} ({\lambda_1} {\tau})^{1-{p} -{q} }
}{{\lambda_1}^2({p} +{q} -1)}
+\frac{{\lambda_2}^2 ({\lambda_1} {\tau})^{1-2 {p} }
}{2{\lambda_1}^2(2 {p} -1)}
+\frac{{\lambda_3}^2 ({\lambda_1} {\tau})^{1-2 {q} }
}{2{\lambda_1}^2(2 {q} -1)}
, \end{equation} 
\begin{equation} 
\label{GeneralX2}
x^2 ( {\tau} )  = 
\frac{1}{{\lambda_1}}
\left(
\frac{{\lambda_2}}{1-2 {p} }
({\lambda_1} {\tau})^{1-2 {p} } 
-\frac{{\cos{({\alpha})}}\, {\lambda_3}}{{p} +{q} -1}
 ({\lambda_1} {\tau})^{1-{p} -{q} }
\right) 
, \end{equation} 
\begin{equation} 
\label{GeneralX3}
x^3 ( {\tau} )  = 
\frac{1}{{\lambda_1}}
\left(
\frac{{\lambda_3}}{1-2 {q} }
({\lambda_1} {\tau})^{1-2 {q} } 
-\frac{{\cos{({\alpha})}}\, {\lambda_2} }{{p} +{q} -1}
 ({\lambda_1} {\tau})^{1-{q} -{p} }
\right) 
.\end{equation} 

We will use the parametrization of solutions of the Einstein equations by two independent angular parameters ${\alpha}$ and ${\beta}$:
$$
{{p}}= \frac{1}{2} \, \bigl(1+\sin {\alpha} \sin {\beta}+\cos {\beta}\bigr)
,\qquad
{{q}}= \frac{1}{2} \, \bigl(1-\sin {\alpha} \sin   {\beta}+\cos {\beta}\bigr)
,$$
$$
0<{\alpha} <\pi
,\qquad
0<{{\beta}}<\pi
.$$
Solving the system of equations (\ref{Deviation1})-(\ref{Deviation2}) for a gravitational wave (\ref{Metric3}), we obtain the geodesic deviation vector $\eta^{\gamma}(\tau)$ in the privileged coordinate system in the following form:

\begin{equation} 
{x^0}(\tau)={\lambda_1}\tau
, \end{equation}
\begin{equation} 
 \eta^0(\tau) ={\rho_1} \tau
 -{\lambda_1} {\Omega} 
  ,\qquad 
 \Omega=\vartheta_1 \lambda_1+\vartheta_2 \lambda_2+\vartheta_3 \lambda_3
, \end{equation} 
$$
 \eta^1(\tau) = 
 \frac{(\lambda_2\rho_1-\lambda_1\rho_2) 
 \left[
{\lambda_2}
  -
 {\lambda_3} \cos {\alpha} {\left(x^0\right)}^{\sin {\alpha} \sin {\beta}} 
 (1+\sin {\alpha} \tan {\beta})
 \right]
 }{{\lambda_1}^3 \left(\cos {\beta}+\sin {\alpha} \sin{\beta}\right)
  {\left(x^0\right)}^{\sin {\alpha} \sin {\beta}+\cos {\beta}} 
 }
 $$
 $$
\mbox{}
 +\frac{(\lambda_3\rho_1-\lambda_1\rho_3) 
\left[
{\lambda_2} \cos {\alpha} (\sin {\alpha} \tan {\beta}-1)+{\lambda_3} {\left(x^0\right)}^{\sin {\alpha} \sin {\beta}}
\right]
}{{\lambda_1}^3 
\left(\cos {\beta}-\sin {\alpha} \sin {\beta}\right)
 {\left(x^0\right)}^{\cos {\beta}}
}
 $$
 $$
 \mbox{}
- \frac{
 {\left(x^0\right)}^{-\sin {\alpha} \sin {\beta}-\cos {\beta}-1} 
  }{2 {\lambda_1}^3 
  } 
\biggl[
 -2 {\lambda_1}^3 {\vartheta_1} {\left(x^0\right)}^{\sin {\alpha} \sin {\beta}+\cos {\beta}+1}
 $$
 $$
\mbox{}
 +{\lambda_1}^2 {\Omega} 
 \left({\lambda_2}^2+{\lambda_3}^2 {\left(x^0\right)}^{2 \sin {\alpha} \sin {\beta}}+{\left(x^0\right)}^{\sin {\alpha} \sin {\beta}+\cos {\beta}+1}
 \right)
 $$
 $$
\mbox{}
 -2 {\lambda_2} {\lambda_3} \cos {\alpha} 
 \left({\lambda_1}^2 {\Omega}-{\rho_1} {x^0}
 \right) {\left(x^0\right)}^{\sin {\alpha} \sin {\beta}}
 $$
\begin{equation} 
\mbox{}
 +{\rho_1} {x^0} 
 \left(-{\lambda_2}^2-{\lambda_3}^2 {\left(x^0\right)}^{2 \sin {\alpha} \sin {\beta}}+{\left(x^0\right)}^{\sin {\alpha} \sin {\beta}+\cos {\beta}+1}
 \right)
\biggr]
, \end{equation} 
$$
 \eta^2(\tau) = 
 \frac{
 \left({\lambda_1}^2 {\Omega}-{\rho_1} {x^0}\right) 
 \left({\lambda_2}-{\lambda_3} \cos {\alpha} {\left(x^0\right)}^{\sin {\alpha} \sin {\beta}}\right)
 }{{\lambda_1}^2 
 \,{\left(x^0\right)}^{1+\sin {\alpha} \sin {\beta}+\cos {\beta}} 
 }
 $$
$$
 \mbox{}
 -\frac{(\lambda_2\rho_1-\lambda_1\rho_2) {\left(x^0\right)}^{-\sin {\alpha} \sin {\beta}-\cos {\beta}}}{{\lambda_1}^2 (\sin {\alpha} \sin {\beta}+\cos {\beta})}
 $$
 \begin{equation} 
  \mbox{}
 +\frac{(\lambda_3\rho_1-\lambda_1\rho_3) \cos {\alpha} 
 {\left(x^0\right)}^{-\cos {\beta}}}{{\lambda_1}^2\cos{\beta} }+{\vartheta_2} 
, \end{equation} 
$$
 \eta^3(\tau) = \frac{{\left(x^0\right)}^{-\cos {\beta}-1} \left({\lambda_1}^2 {\Omega}-{\rho_1} {x^0}\right) \left({\lambda_3} {\left(x^0\right)}^{\sin {\alpha} \sin {\beta}}-{\lambda_2} \cos {\alpha}\right)}{{\lambda_1}^2}
 $$
\begin{equation} 
\mbox{}
 +\frac{(\lambda_2\rho_1-\lambda_1\rho_2) \cos {\alpha}
 {\left(x^0\right)}^{-\cos {\beta}}
 }{{\lambda_1}^2\cos{\beta}}
  -\frac{(\lambda_3\rho_1-\lambda_1\rho_3) {\left(x^0\right)}^{\sin {\alpha} \sin {\beta}-\cos {\beta}}}{{\lambda_1}^2 (\cos {\beta}-\sin {\alpha} \sin {\beta})}+{\vartheta_3} 
, \end{equation} 
where $\tau$ is the proper time on the base geodesic line, the angular parameters 
${\alpha}$ and ${\beta}$ are the parameters of the gravitational model, the parameters $\lambda_{k}$, $\rho_{k}$ and $\vartheta_{k}$ (${k}=1,2,3$) are independent parameters, determined by the initial conditions on the geodesics.
 
Then the tidal acceleration $A^{\gamma} (\tau)$ in a gravitational wave (\ref{Metric3}) in the privileged coordinate system takes the following form:

\begin{equation}
A^{\gamma} (\tau) = 
\frac{D^2 }{{d\tau}^2}\,\eta^{\gamma}(\tau)
,\qquad
x^0=\lambda_1\tau
,\end{equation}
\begin{equation}
A^0 = 0
,\end{equation}
$$
A^1 (\tau) = 
\frac{
{\lambda_1}
{\vartheta_2} 
\sin {\beta}
}{2}
\left(
\frac{
{\lambda_3}   (\cos {\alpha} \sin {\beta}+\cot {\alpha} \cos {\beta})
}{{\left(x^0\right)}^{2-\sin {\alpha} \sin {\beta}}}
-\frac{
{\lambda_2} \cos {\beta}}{ {\left(x^0\right)}^2 \sin {\alpha} }\right)
$$
$$
\mbox{}
+
\frac{
{\lambda_1} 
{\vartheta_3} 
 \sin {\beta}
}{2}
\left(  
\frac{
{\lambda_2}  (\cos {\alpha} \sin {\beta}-\cot {\alpha} \cos {\beta})
}{ {\left(x^0\right)}^{2+\sin {\alpha} \sin {\beta}}}
+\frac{{\lambda_3}\cos {\beta}}{ {\left(x^0\right)}^2  \sin {\alpha} }\right)
$$
$$
\mbox{}
-\frac{{\lambda_2} (\lambda_2\rho_1-\lambda_1\rho_2) 
\left[\cos 2 ({\alpha}-{\beta})+\cos 2 ({\alpha}+{\beta})-2 \cos 2 {\alpha}-2 \cos 2 {\beta}-6\right]
}{16 {\lambda_1} 
\csc {\alpha} \cot {\beta} 
(\sin {\alpha} \sin {\beta}+\cos {\beta})
{\left(x^0\right)}^{2+\sin {\alpha} \sin {\beta}+\cos {\beta}}
}
$$
\begin{equation}
\mbox{}
+\frac{{\lambda_3} (\lambda_3\rho_1-\lambda_1\rho_3) 
\left[\cos 2 ({\alpha}-{\beta})+\cos 2 ({\alpha}+{\beta})-2 \cos 2 {\alpha}-2 \cos 2 {\beta}-6\right]
}{16 {\lambda_1} 
\csc {\alpha} \cot {\beta} 
(\cos {\beta}-\sin {\alpha} \sin {\beta})
{\left(x^0\right)}^{2-\sin {\alpha} \sin {\beta}+\cos {\beta}}
}
,\end{equation}

$$
A^2 (\tau)  = \frac{{\lambda_1}^2 {\vartheta_2} \csc {\alpha} \sin {\beta} \cos {\beta}}{2 {\left(x^0\right)}^2}
-\frac{ 
{\lambda_1}^2 {\vartheta_3} 
\sin {\beta}  (\sin {\alpha} \sin {\beta}-\cos {\beta})
}{2\tan {\alpha} \,{\left(x^0\right)}^{2+\sin {\alpha} \sin {\beta}}
}
$$
\begin{equation}
\mbox{}
+\frac{
(\lambda_2\rho_1-\lambda_1\rho_2)
\left[
\cos 2 ({\alpha}-{\beta})+\cos 2 ({\alpha}+{\beta})-2 \cos 2 {\alpha}-2 \cos 2 {\beta}-6
\right]
}{16
\csc {\alpha} \cot {\beta} 
 (\sin {\alpha} \sin {\beta}+\cos {\beta})
{\left(x^0\right)}^{2+\sin {\alpha} \sin {\beta}+\cos {\beta}}
}
,\end{equation}

$$
A^3 (\tau)  = -\frac{{\lambda_1}^2 {\vartheta_2} \sin {\beta} 
 \left(\sin {\alpha} \cos {\alpha} \sin ^2{\beta}-\cot {\alpha} \cos ^2{\beta}\right)
 }{2 (\sin {\alpha} \sin {\beta}-\cos {\beta}) {\left(x^0\right)}^{2-\sin {\alpha} \sin {\beta}}
 }
-\frac{{\lambda_1}^2 {\vartheta_3} 
\sin (2 {\beta})}{4 {\left(x^0\right)}^2 \sin{\alpha}}
$$
\begin{equation}
\mbox{}
-\frac{(\lambda_3\rho_1-\lambda_1\rho_3) 
\left[
\cos 2 ({\alpha}-{\beta})+\cos 2 ({\alpha}+{\beta})-2 \cos 2 {\alpha}-2 \cos 2 {\beta}-6
\right]
}{16 
\cot {\beta} 
(\csc {\alpha} \cos {\beta}-\sin {\beta})
 {\left(x^0\right)}^{2-\sin {\alpha} \sin {\beta}+\cos {\beta}}
}
,\end{equation}
where $\tau$ is the proper time of the test particle on the base geodesic line, the angular parameters 
${\alpha}$ and ${\beta}$ 
are the parameters of the gravitational wave, the parameters 
$\lambda_{k}$, $\rho_{k}$ and $\vartheta_{k}$ (${k}=1,2,3$) 
are independent parameters, determined by the initial conditions on the geodesics (initial or boundary values of coordinates and momenta of test particles on adjacent geodesic lines).

Thus, in the privileged coordinate system, we have obtained the exact form of the vector of deviation of geodesic and tidal acceleration of test particles.
Using these results, it is possible to calculate the physical effects from the gravitational wave, including the calculation of the radiation of charges according to their given motion (deviation vector). Unfortunately, the privileged coordinate system does not give the usual physical picture, since the time and spatial coordinates are mixed here.

\section{The synchronous coordinate system}

The advantage of synchronous coordinate systems is that in these coordinate systems time and spatial coordinates are separated and time synchronization is allowed at different points in space (see~\cite{LandauEng1}), which is physically clear and convenient for real physical measurements (including precise laser measurements).

The resulting form of the trajectories of test particles in the gravitational wave 
(\ref{GeneralX0})-(\ref{GeneralX3}) makes it possible to carry out
transition from the privileged coordinate system to the synchronous coordinate system ${\tilde x}{}^{\gamma}$, where the test particle is at rest on the base geodesic (the system of a freely falling observer). 

The ability to analytically construct a transition to a synchronous coordinate system is based on the fact that in our model we have found a complete integral for the Hamilton-Jacobi equation of test particles.
The transition to the synchronous coordinate system ${\tilde x}{}^{\gamma}$ can be explicitly implemented using the relations (\ref{GeneralX0})-(\ref{GeneralX3}) according to the rules
 (see~\cite{LandauEng1}):

\begin{equation}
x^{\gamma} \to {\tilde x}{}^{\gamma}=\left(\tau,\lambda_1,\lambda_2,\lambda_3 \right)
.\end{equation}
In the synchronous (laboratory) coordinate system thus obtained, the test particle (freely falling observer) will be at rest, and the observer's proper time $\tau$ will be a time variable.

The formulas for the transition from the privileged coordinate system $x^{\gamma}$ to the synchronous coordinate system $\tilde x{}^{\gamma}$, taking into account the type of test particle trajectories (\ref{GeneralX0})-(\ref{GeneralX3}) obtained earlier, into the angular parameterization will take the following form:

\begin{equation} 
x^0 = {\tilde x{}^1} {\tau} 
, \end{equation} 
$$
x^1 = 
-\frac{\cos {\beta} ({\tilde x{}^1} {\tau})^{-\sin {\alpha} \sin {\beta}-\cos {\beta}} 
}{2 {\tilde x{}^1}^2 \left(\cos ^2{\beta}-\sin ^2{\alpha} \sin ^2{\beta}\right)} 
\biggl[
{\tilde x{}^2}^2( 1-\sin {\alpha} \tan {\beta})
$$
$$
\mbox{}
+2  \cos {\alpha} \left(\sin ^2{\alpha} \tan ^2{\beta}-1\right)
{\tilde x{}^2} {\tilde x{}^3} ({\tilde x{}^1} {\tau})^{\sin {\alpha} \sin {\beta}}
$$
$$
\mbox{}
+{\tilde x{}^3}^2 (1+ \sin {\alpha} \tan {\beta}) ({\tilde x{}^1} {\tau})^{2 \sin {\alpha} \sin {\beta}}
$$
\begin{equation} 
\mbox{}
+\cos {\beta}\left(\sin ^2{\alpha} \tan^2 {\beta}-1 \right) 
({\tilde x{}^1} {\tau})^{1+\sin {\alpha} \sin {\beta}+\cos {\beta}}
\biggr]
, \end{equation} 
\begin{equation} 
x^2 = \frac{{\tau}
\left[
{\tilde x{}^2}-{\tilde x{}^3} \cos {\alpha} (\sin {\alpha} \tan {\beta}+1) ({\tilde x{}^1} {\tau})^{\sin {\alpha} \sin {\beta}}
\right]
}{(\sin {\alpha} \sin {\beta}+\cos {\beta})
({\tilde x{}^1} {\tau})^{1+\sin {\alpha} \sin {\beta}+\cos {\beta}} 
} 
, \end{equation} 
\begin{equation} 
x^3 = \frac{
\tau
\left[
{\tilde x{}^3} ({\tilde x{}^1} {\tau})^{\sin {\alpha} \sin {\beta}}+{\tilde x{}^2} \cos {\alpha} (\sin {\alpha} \tan {\beta}-1)
\right]
}{ 
(\cos {\beta}-\sin {\alpha} \sin {\beta})
({\tilde x{}^1} {\tau})^{1+\cos {\beta}} 
} 
. \end{equation} 
Under these transformations, the 4-vector of the test particle velocity on the base geodesic line in the synchronous coordinate system ${\tilde x}{}^{\gamma}$ takes the form
$
\tilde u{}^{\gamma}=\left\{ 1,\, 0,\, 0,\, 0\right\}
$,
those the test particle is at rest on the base geodesic line in the chosen synchronous coordinate system.
Moreover, the constant spatial coordinates of the test particle on the base geodesic line are given by the  parameters $\lambda_1$, $\lambda_2$, and $\lambda_3$, which determines the physical meaning of these constants in the synchronous coordinate system.

The gravitational wave metric (\ref{Metric3}) in the synchronous coordinate system 
${\tilde x}{}^{\gamma}$ takes the following form
($\tau$ is a new time variable)
\begin{equation} 
\label{SynchrMetric4AA}
{ds}^2=\tilde g{}_{\alpha\beta}\,{d\tilde x{}^\alpha}{d\tilde x{}^\beta}
={d\tau}^2-{dl}^2,
\qquad
{dl}^2=-\tilde g{}_{{i}{j}}(\tau,\tilde x{}^{k}) \,
{d\tilde x{}^{i}}{d\tilde x{}^{j}}
, \end{equation} 
$$
{i},{j},{k} = 1\ldots 3;
\qquad
{\alpha},{\beta},{\gamma}=0\ldots 3,
$$
where the speed of light is chosen to be unity, $l$ is the spatial distance, and the metric components take the form
\begin{equation} 
\label{SynchrMetric4A}
\tilde g{}^{00} = 1 
,\qquad
\tilde g{}^{01} = 
\tilde g{}^{02} = 
\tilde g{}^{03} = 0 
, \end{equation} 
\begin{equation} 
\label{SynchrMetric4B}
\tilde g{}^{1k} = -\frac{{\tilde x{}^1}{\tilde x{}^k} }{{\tau}^2} 
%
, \end{equation} 
\begin{equation} 
\label{SynchrMetric4C}
\tilde g{}^{22} = -\frac{ \cos ^2{\beta} (\csc {\alpha} \cos {\beta}+\sin {\beta})^2 
\,({\tilde x{}^1})^2
}{ 
\left(\cos ^2{\alpha} \sin ^2{\beta}+\cos ^2{\beta}\right)
({\tilde x{}^1} {\tau})^{1-\sin {\alpha} \sin {\beta}-\cos {\beta}}
}
-\frac{({\tilde x{}^2})^2}{{\tau}^2} 
, \end{equation} 
\begin{equation} 
\label{SynchrMetric4D}
\tilde g{}^{23} = -\frac{
\cos {\alpha} \cos ^2{\beta} 
\left(\csc^2 {\alpha} \cos^2 {\beta}-\sin^2 {\beta}\right) 
\,({\tilde x{}^1})^2 
}{
\left(\cos ^2{\alpha} \sin ^2{\beta}+\cos ^2{\beta}\right)
 ({\tilde x{}^1} {\tau})^{1-\cos {\beta}}
}
-\frac{{\tilde x{}^2} {\tilde x{}^3}}{{\tau}^2} 
, \end{equation} 
\begin{equation} 
\label{SynchrMetric4E}
\tilde g{}^{33} = -\frac{
 \cos ^2{\beta} (\sin {\beta}-\csc {\alpha} \cos {\beta})^2
\, ({\tilde x{}^1})^2
}{
\left(\cos ^2{\alpha} \sin ^2{\beta}+\cos ^2{\beta}\right)
({\tilde x{}^1} {\tau})^{1+\sin {\alpha} \sin {\beta}-\cos {\beta}}
}
-\frac{({\tilde x{}^3})^2}{{\tau}^2} 
. \end{equation} 
The chosen synchronous coordinate system has a singularity at the origin.

Geodesic deviation vector $\tilde\eta{}^{\gamma}(\tau)$ in a gravitational wave with the metric (\ref{SynchrMetric4A})-(\ref{SynchrMetric4E}) in the synchronous coordinate system ${\tilde x}{}^ k$ has only spatial components and takes the following form:

\begin{equation} 
\tilde\eta{}^0= 0 
, \end{equation} 
\begin{equation} 
\tilde\eta{}^1(\tau)= {\rho_1}-\frac{{\lambda_1} {\Omega}}{{\tau}} 
  ,\qquad 
 \Omega=\vartheta_1 \lambda_1+\vartheta_2 \lambda_2+\vartheta_3 \lambda_3
, \end{equation} 
$$
\tilde\eta{}^2(\tau)= 
{\lambda_1} {\vartheta_2} 
\,
\frac{
\cos ^2{\beta} 
\,
(\sin {\alpha} \sin {\beta}+\cos {\beta})
 }{ 
\sin ^2{\alpha} 
\,
(\cos ^2{\alpha} \sin ^2{\beta}+\cos ^2{\beta}) }
\,
({\lambda_1} {\tau})^{\sin {\alpha} \sin {\beta}+\cos {\beta}}
$$
\begin{equation}
\mbox{}
+
{\lambda_1} {\vartheta_3} 
\,
\frac{
 \cos {\beta} 
\left(\cot {\alpha} \csc {\alpha} \cos ^2{\beta}-\cos {\alpha} \sin ^2{\beta}\right)}{\cos ^2{\alpha} \sin ^2{\beta}+\cos ^2{\beta}}
\,
({\lambda_1} {\tau})^{\cos {\beta}} 
\mbox{}
-\frac{{\lambda_2} {\Omega}}{{\tau}}+{\rho_2} 
, \end{equation} 
$$
\tilde\eta{}^3(\tau)= 
{\lambda_1} {\vartheta_2} 
\,
\frac{
\cos {\beta}
 \left(\cot {\alpha} \csc {\alpha} \cos ^2{\beta}-\cos {\alpha} \sin ^2{\beta}\right)}{\cos ^2{\alpha} \sin ^2{\beta}+\cos ^2{\beta}}
 \,
  ({\lambda_1} {\tau})^{\cos {\beta}} 
$$
\begin{equation} 
\mbox{}
+
{\lambda_1} {\vartheta_3} 
\,
\frac{
\cos ^2{\beta} (\cos {\beta}-\sin {\alpha} \sin {\beta}) 
}{ \sin ^2{\alpha} (\cos ^2{\alpha} \sin ^2{\beta}+\cos ^2{\beta})}
\,
({\lambda_1} {\tau})^{\cos {\beta}-\sin {\alpha} \sin {\beta}}
-\frac{{\lambda_3} {\Omega}}{{\tau}}+{\rho_3} 
,\end{equation} 
where $\tau$ is the proper time on the base geodesic line, the parameters ${\alpha}$ and ${\beta}$ are the angular parameters of the gravitational wave, the parameters 
$\lambda_{k}$, $\rho_{k}$ and $\vartheta_{k}$ (${k}=1,2,3$) are independent parameters, determined by the initial conditions on the geodesics.
 
The tidal acceleration of test particles in a gravitational wave (\ref{SynchrMetric4A})-(\ref{SynchrMetric4E})  in a synchronous coordinate system takes the following simple form:

$$
\tilde A{}^{\gamma} (\tau) = 
\frac{D^2 }{{d\tau}^2}\,\tilde \eta{}^{\gamma}(\tau)
,$$
\begin{equation} 
\tilde A{}^0= 0 
,\qquad
\tilde A{}^1= 0 
, \end{equation} 
$$
\tilde A{}^2 (\tau) = -\frac{(\lambda_2\rho_1-\lambda_1\rho_2) \csc {\alpha} \sin {\beta} \cos {\beta}}{2 {\lambda_1} {\tau}^2}
$$
$$
\mbox{}
+\frac{{\lambda_1}^3 {\vartheta_2} \csc {\alpha} \sin (2 {\beta}) (\sin {\alpha} \sin {\beta}+\cos {\beta}) 
}{4 
({\lambda_1} {\tau})^{2+\sin {\alpha} \sin {\beta}+\cos {\beta}}}
$$
\begin{equation} 
\mbox{}
+\frac{
 {\lambda_1} (\lambda_3\rho_1-\lambda_1\rho_3) 
 \sin {\beta} (\sin {\alpha} \sin {\beta}+\cos {\beta}) 
 }{2 \tan {\alpha}({\lambda_1} {\tau})^{2-\sin {\alpha} \sin {\beta}}  }
, \end{equation} 
$$
\tilde A{}^3 (\tau) = \frac{(\lambda_3\rho_1-\lambda_1\rho_3) \csc {\alpha} \sin {\beta} \cos {\beta}}{2 {\lambda_1} {\tau}^2}
$$
$$
\mbox{}
+\frac{{\lambda_1}^3 {\vartheta_3} \sin {\beta} \cos {\beta} (\sin {\beta}-\csc {\alpha} \cos {\beta})
 }{2 
 ({\lambda_1} {\tau})^{2-\cos {\beta}+\sin {\alpha} \sin {\beta}}}
$$
\begin{equation} 
\mbox{}
+\frac{
 {\lambda_1} (\lambda_2\rho_1-\lambda_1\rho_2) \sin {\beta}
  (\cos {\alpha} \sin {\beta}-\cot {\alpha} \cos {\beta}) 
 }{2({\lambda_1} {\tau})^{2+\sin {\alpha} \sin {\beta}} }
. \end{equation} 
Thus, tidal accelerations in a gravitational wave (\ref{Metric3}) in the synchronous coordinate system arise only in the plane of variables $\tilde x{}^2$ and $\tilde x{}^3$, and the gravitational wave propagates along coordinates $\tilde x{}^1$. The time variable $\tau$ is the proper time of the observer freely falling along the base geodetic line, with which the selected synchronous coordinate system is associated.

The results obtained here in the synchronous coordinate system now make it possible to explicitly write the Maxwell equations for calculating the radiation of a charge moving in the field of a gravitational wave according to the obtained deviation vector.

Obtaining an exact model of radiation, similar to the Lienard-Wiechert radiation for the problem under consideration, as well as for spaces of other types of Bianchi, will also allow us to estimate the presence of certain types of symmetry in the early stages of the dynamics of the universe.

We intend to consider such problems in our future studies.

\section{Special case I of integration of the Hamilton-Jacobi equation (${p}+{q}=1$)}

When integrating the Hamilton-Jacobi equation of test particles for the \mbox{Shapovalov} type III space in the Bianchi type VI cosmological model for the metric (\ref{Metric}), a special case arises when ${p}+{q}=1$.
Let us consider this case and set for shortening the notation

\begin{equation} 
{p}=\frac{1}{2}+\omega
,\qquad
{q}=\frac{1}{2}-\omega
.\end{equation} 
Solving the Einstein vacuum equations in this case gives us:

\begin{equation} 
\label{MetricPrivIA}
 {ds}^2=dx^0dx^1-\frac{1}{4\omega^2}\left[
 {\left(x^0\right)}^{1+2\omega}{(dx^2)}^2
 +2{a}{x^0}{dx^2}{dx^3}
 + {\left(x^0\right)}^{1-2\omega}{(dx^3)}^2
 \right]
,
\end{equation} 
\begin{equation} 
\label{MetricPrivIB}
\Lambda=0
  ,\qquad
 0<\omega^2 \le \frac{1}{4}
  ,\qquad
  \omega\ne 0
   ,\qquad
  {a}^2=1-4\omega^2
,\end{equation} 
where $x^0$ is a wave variable, $\omega$ is a constant parameter.

The nonzero components of the Riemann curvature tensor take the form

\begin{equation} 
R_{0202} =- \frac{\left(1-4 \omega ^2\right) {\left(x^0\right)}^{2 \omega -1}}{8 \omega ^2}
,\end{equation} 
\begin{equation} 
R_{0302} =-\frac{a}{8 \omega ^2 {x^0}}
,\end{equation} 
\begin{equation} 
R_{0303} = -\frac{\left(1-4 \omega ^2\right) {\left(x^0\right)}^{-2 \omega -1}}{8 \omega ^2}
.\end{equation} 
From Einstein equations we get: if  ${a}=0$ ($\omega=\pm1/2$), then not only the Einstein tensor but the Riemann tensor $R_{{\alpha}{\beta}{\gamma}{\delta}}$ is null, and spacetime is flat.

The solution of the Hamilton-Jacobi equation for a test particle in a privileged coordinate system in this case leads to the following form of the test particle trajectory equations ($\tau$ is the proper time of the particle):

\begin{equation} 
\label{MuPlusNuX0}
x^0(\tau) = {\lambda_1} {\tau} 
, \end{equation} 
\begin{equation} 
\label{MuPlusNuX1}
x^1(\tau) = \frac{-4 {a} {\lambda_2} {\lambda_3} \omega  \log ({\lambda_1} {\tau})-{\lambda_2}^2 ({\lambda_1} {\tau})^{-2 \omega }+{\lambda_3}^2 ({\lambda_1} {\tau})^{2 \omega }+2 {\lambda_1} \omega  {\tau}}{4 {\lambda_1}^2 \omega } 
, \end{equation} 
\begin{equation} 
\label{MuPlusNuX2}
x^2(\tau) = \frac{2 {a} {\lambda_3}{\omega } \log ({\lambda_1} {\tau})
+
{\lambda_2} ({\lambda_1} {\tau})^{-2 \omega }
}{2 {\lambda_1}{\omega }} 
, \end{equation} 
\begin{equation} 
\label{MuPlusNuX3}
x^3(\tau) = -\frac{{\lambda_3} ({\lambda_1} {\tau})^{2 \omega }-2 {a} {\lambda_2} \omega  \log ({\lambda_1} {\tau})}{2 {\lambda_1} \omega } 
, \end{equation} 
where the parameters $\lambda_{i}$ are the integrals of particle motion determined by the initial conditions.

Solution for the geodesic deviation vector $\eta^{\alpha}(\tau)$ for a gravitational wave (\ref{MetricPrivIA})-(\ref{MetricPrivIB}) in a privileged coordinate system, taking into account the type of test particle trajectories (\ref{MuPlusNuX0})-(\ref{MuPlusNuX3}) we find from the equations (\ref{Deviation1})-(\ref{Deviation2}) in the following form

\begin{equation} 
\label{MuPlusNuEta0}
 \eta^0(\tau) = {\rho_1} {\tau}-{\lambda_1} {\Omega} 
  ,\qquad 
 \Omega=\vartheta_1 \lambda_1+\vartheta_2 \lambda_2+\vartheta_3 \lambda_3
 ,\qquad 
  x^0 = {\lambda_1} {\tau} 
, \end{equation} 
$$
 \eta^1(\tau) = \frac{{\rho_1} \left(-2 {a} {\lambda_2} {\lambda_3}+{\lambda_2}^2 {\left(x^0\right)}^{-2 \omega }+{\lambda_3}^2 {\left(x^0\right)}^{2 \omega }-{x^0}\right)}{2 {\lambda_1}^3}
 $$
 $$
 +\frac{(\lambda_2\rho_1-\lambda_1\rho_2) \left(2 {a} {\lambda_3} \log \left(x^0\right)+\frac{{\lambda_2} {\left(x^0\right)}^{-2 \omega }}{\omega }\right)}{2 {\lambda_1}^3}
 $$
 $$
 +\frac{(\lambda_3\rho_1-\lambda_1\rho_3) \left(2 {a} {\lambda_2} \log \left(x^0\right)-\frac{{\lambda_3} {\left(x^0\right)}^{2 \omega }}{\omega }\right)}{2 {\lambda_1}^3}
 $$
\begin{equation}   
\label{MuPlusNuEta1}
 -\frac{{\Omega} \left(-2 {a} {\lambda_2} {\lambda_3}+{\lambda_2}^2 {\left(x^0\right)}^{-2 \omega }+{\lambda_3}^2 {\left(x^0\right)}^{2 \omega }+{x^0}\right)}{2 {\lambda_1} {x^0}}+{\vartheta_1} 
, \end{equation} 
$$
 \eta^2(\tau) = \frac{{\left(x^0\right)}^{-2 \omega -1} \left({\lambda_1}^2 {\Omega}-{\rho_1} {x^0}\right) \left({\lambda_2}-{a} {\lambda_3} {\left(x^0\right)}^{2 \omega }\right)}{{\lambda_1}^2}
 $$
\begin{equation} 
\label{MuPlusNuEta2}
 -\frac{{a} (\lambda_3\rho_1-\lambda_1\rho_3) \log \left(x^0\right)}{{\lambda_1}^2}-\frac{(\lambda_2\rho_1-\lambda_1\rho_2) {\left(x^0\right)}^{-2 \omega }}{2 {\lambda_1}^2 \omega }+{\vartheta_2} 
, \end{equation} 
$$
 \eta^3(\tau) = -\frac{\left({\lambda_1}^2 {\Omega}-{\rho_1} {x^0}\right) \left({a} {\lambda_2}-{\lambda_3} {\left(x^0\right)}^{2 \omega }\right)}{{\lambda_1}^2 {x^0}}
 $$
\begin{equation} 
\label{MuPlusNuEta3}
 -\frac{{a} (\lambda_2\rho_1-\lambda_1\rho_2) \log \left(x^0\right)}{{\lambda_1}^2}+\frac{(\lambda_3\rho_1-\lambda_1\rho_3) {\left(x^0\right)}^{2 \omega }}{2 {\lambda_1}^2 \omega }+{\vartheta_3} 
, \end{equation} 
where $\omega$ and ${a}$ are the parameters of the considered cosmological wave model of spacetime determined by relations (\ref{MetricPrivIB}),
$\lambda_{i}$, $\vartheta_{i}$ and $\rho_{i}$ are independent parameters of geodesic lines determined from the initial conditions, $\tau$ is proper time of a test particle on the base geodesic line.

The tidal acceleration $A^{\gamma}$ of a gravitational wave (\ref{MetricPrivIA})-(\ref{MetricPrivIB}) in the privileged coordinate system takes the form

$$
A^{\gamma}(\tau)=\frac{D^2 \eta^{\gamma}}{{d\tau}^2}
,\qquad
x^0 = {\lambda_1} {\tau} 
,\qquad
{a}^2=1-4\omega^2
, $$
\begin{equation} 
A^0 = 0
,\end{equation} 
 $$
A^1\left(x^0\right) =
 (\lambda_2\rho_1-\lambda_1\rho_2) \left(-\frac{ (1-4\omega^2) {\lambda_2} {\left(x^0\right)}^{-2 (\omega +1)} \log \left(x^0\right)}{2 {\lambda_1}}-\frac{{a} {\lambda_3}}{4 {\lambda_1} \omega  {\left(x^0\right)}^2}\right)
 $$
 $$
 +(\lambda_3\rho_1-\lambda_1\rho_3) \left(\frac{{a} {\lambda_2}}{4 {\lambda_1} \omega  {\left(x^0\right)}^2}-\frac{ (1-4\omega^2){\lambda_3} {\left(x^0\right)}^{2 \omega -2} \log \left(x^0\right)}{2 {\lambda_1}}\right)
 $$
\begin{equation} 
 +\frac{1}{2} {a} {\lambda_1} {\lambda_2} {\vartheta_3} {\left(x^0\right)}^{-2 (\omega +1)}+\frac{1}{2} {a} {\lambda_1} {\lambda_3} {\vartheta_2} {\left(x^0\right)}^{2( \omega -1)}
,\end{equation} 
$$
A^2 \left(x^0\right)= 
\frac{1}{2} \left(1-4 \omega ^2\right) (\lambda_2\rho_1-\lambda_1\rho_2) {\left(x^0\right)}^{-2 (\omega +1)} \log \left(x^0\right)
$$
\begin{equation} 
-\frac{1}{2} {a} {\lambda_1}^2 {\vartheta_3} {\left(x^0\right)}^{-2 (\omega +1)}-\frac{{a} (\lambda_3\rho_1-\lambda_1\rho_3)}{4 \omega  {\left(x^0\right)}^2}
,\end{equation} 
$$
A^3\left(x^0\right) =
\frac{1}{2} \left(1-4 \omega ^2\right) (\lambda_3\rho_1-\lambda_1\rho_3) {\left(x^0\right)}^{2 \omega -2} \log \left(x^0\right)
$$
\begin{equation} 
 -\frac{1}{2} {a} {\lambda_1}^2 {\vartheta_2} {\left(x^0\right)}^{2( \omega -1)}
 +\frac{{a} (\lambda_2\rho_1-\lambda_1\rho_2)}{4 \omega  {\left(x^0\right)}^2}
,\end{equation} 
where $\omega$ 
is the parameter
of the cosmological wave model of spacetime determined by relations (\ref{MetricPrivIB}),
$\lambda_{i}$, $\vartheta_{i}$ and $\rho_{i}$ are parameters of geodesic lines determined from the initial conditions, $\tau$ is proper time of a test particle on the base geodesic line.

The transition to the synchronous coordinate system ${\tilde x}{}^{\alpha}$, where the test particle is at rest on the base geodesic (freely falling observer) can be done using the relations (\ref{MuPlusNuX0})-(\ref{MuPlusNuX3}) :

\begin{equation}
x^{\alpha} \to {\tilde x}{}^{\alpha}=\left(\tau,\lambda_1,\lambda_2,\lambda_3 \right).
\end{equation}

The gravitational wave metric (\ref{MetricPrivIA})-(\ref{MetricPrivIB}) in the synchronous coordinate system takes the form

$$
\tilde g{}^{00} = 1 
,\quad
\tilde g{}^{01} = 0 
,\quad
\tilde g{}^{02} = 0 
,\quad
\tilde g{}^{03} = 0 
, $$
$$
\tilde g{}^{11} = -\frac{{\tilde x{}^1}^2}{{\tau}^2} 
,\quad
\tilde g{}^{12} = -\frac{{\tilde x{}^1} {\tilde x{}^2}}{{\tau}^2} 
,\quad
\tilde g{}^{13} = -\frac{{\tilde x{}^1} {\tilde x{}^3}}{{\tau}^2} 
, $$
$$
\tilde g{}^{22} =\frac{ {\tilde x{}^1} \omega ^2 ({\tilde x{}^1} {\tau})^{2 \omega } 
\left[
\omega ^2 \left(4 \omega ^2-1\right) \log ^2({\tilde x{}^1} {\tau})+ \omega  \left(1-4 \omega ^2\right) \log ({\tilde x{}^1} {\tau})-1/4
\right]
}{{\tau}
\left[
\omega ^2 \left(1-4 \omega ^2\right) \log ^2({\tilde x{}^1} {\tau})+1/4\right]^2}
-\frac{{\tilde x{}^2}^2}{{\tau}^2} 
, $$
$$
\tilde g{}^{23} = \frac{{a} {\tilde x{}^1} \omega ^2 
\left[
\omega ^2 \left(1-4 \omega ^2\right) \log ^2({\tilde x{}^1} {\tau})-1/4
\right]
}{{\tau} 
\left[
 \omega ^2 \left(1-4 \omega ^2\right) \log ^2({\tilde x{}^1} {\tau})+1/4
\right]^2}
-\frac{{\tilde x{}^2} {\tilde x{}^3}}{{\tau}^2} 
, $$
$$
\tilde g{}^{33} =\frac{
 {\tilde x{}^1} \omega ^2 ({\tilde x{}^1} {\tau})^{-2 \omega } 
\left[
\omega ^2 \left(4 \omega ^2-1\right) \log ^2({\tilde x{}^1} {\tau})+ \omega  \left(4 \omega ^2-1\right) \log ({\tilde x{}^1} {\tau})-1/4
\right]
 }{{\tau} \left[ \omega ^2 \left(1-4 \omega ^2\right) \log ^2({\tilde x{}^1} {\tau})+1/4\right]^2}
 -\frac{{\tilde x{}^3}^2}{{\tau}^2} 
. $$

The solution for the geodesic deviation vector $\tilde\eta{}^{\gamma}(\tau)$ for a gravitational wave (\ref{MetricPrivIA})-(\ref{MetricPrivIB}) in the synchronous coordinate system can be represented as

\begin{equation} 
\tilde\eta{}^0= 0 
, \end{equation} 
\begin{equation} 
\tilde\eta{}^1(\tau)= {\rho_1}-\frac{{\lambda_1} {\Omega}}{{\tau}} 
,\qquad \Omega=\vartheta_1 \lambda_1+\vartheta_2 \lambda_2+\vartheta_3 \lambda_3
, \end{equation} 
$$
\tilde\eta{}^2(\tau)= \frac{4 {a} {\lambda_1} \omega ^2 {\vartheta_3} \log ({\lambda_1} {\tau})}{4 \omega ^2 \left(1-4 \omega ^2\right) \log ^2({\lambda_1} {\tau})+1}
$$
\begin{equation} 
\mbox{}
+\frac{2 {\lambda_1} \omega  {\vartheta_2} ({\lambda_1} {\tau})^{2 \omega }}{4 \omega ^2 \left(1-4 \omega ^2\right) \log ^2({\lambda_1} {\tau})+1}-\frac{{\lambda_2} {\Omega}}{{\tau}}+{\rho_2} 
, \end{equation} 
$$
\tilde\eta{}^3(\tau)= \frac{4 {a} {\lambda_1} \omega ^2 {\vartheta_2} \log ({\lambda_1} {\tau})}{4 \omega ^2 \left(1-4 \omega ^2\right) \log ^2({\lambda_1} {\tau})+1}
$$
\begin{equation} 
\mbox{}
-\frac{2 {\lambda_1} \omega  {\vartheta_3} ({\lambda_1} {\tau})^{-2 \omega }}{4 \omega ^2 \left(1-4 \omega ^2\right) \log ^2({\lambda_1} {\tau})+1}-\frac{{\lambda_3} {\Omega}}{{\tau}}+{\rho_3} 
, \end{equation} 
where $\omega$ 
is the parameter
of the considered cosmological wave model of spacetime determined by relations (\ref{MetricPrivIB}),
$\lambda_{i}$, $\vartheta_{i}$ and $\rho_{i}$ are parameters of geodesic lines determined from the initial conditions, $\tau$ is proper time of a test particle on the base geodesic line.

The tidal acceleration $\tilde A{}^{\alpha}$ of a gravitational wave (\ref{MetricPrivIA})-(\ref{MetricPrivIB}) in the synchronous coordinate system takes the form
 
$$
\tilde A{}^{\alpha}(\tau)=\frac{D^2 \tilde\eta{}^{\alpha}}{{d\tau}^2}
,\qquad
{a}^2=1-4\omega^2
, $$
\begin{equation} 
\tilde A{}^0= 0 
,\qquad
\tilde A{}^1= 0 
, \end{equation} 
$$
\tilde A{}^2(\tau)= 
\frac{2 (1-4\omega^2){\lambda_1} \omega ^2 {\vartheta_2} ({\lambda_1} {\tau})^{2 \omega } \log ({\lambda_1} {\tau})}{{\tau}^2 \left(4 \omega ^2 \left(4 \omega ^2-1\right) \log ^2({\lambda_1} {\tau})-1\right)}
$$
$$
\mbox{}
+\frac{{a} {\lambda_1} \omega  {\vartheta_3}}{{\tau}^2 \left(4 \omega ^2 \left(4 \omega ^2-1\right) \log ^2({\lambda_1} {\tau})-1\right)}
-\frac{2 (1-4\omega^2) \omega  (\lambda_2\rho_1-\lambda_1\rho_2) \log ({\lambda_1} {\tau})}{{\lambda_1} {\tau}^2 \left(4 \omega ^2 \left(4 \omega ^2-1\right) \log ^2({\lambda_1} {\tau})-1\right)}
$$
\begin{equation} 
\mbox{}
+\frac{{a} {\lambda_1} (\lambda_3\rho_1-\lambda_1\rho_3) ({\lambda_1} {\tau})^{2 (\omega -1)} \left(4 \omega ^2 \left(4 \omega ^2-1\right) \log ^2({\lambda_1} {\tau})+1\right)}{2 \left(4 \omega ^2 \left(4 \omega ^2-1\right) \log ^2({\lambda_1} {\tau})-1\right)} 
, \end{equation} 
$$
\tilde A{}^3(\tau)= 
\frac{2 (1-4\omega^2) {\lambda_1}^3 \omega ^2 {\vartheta_3} ({\lambda_1} {\tau})^{-2 (\omega +1)} \log ({\lambda_1} {\tau})}{4 \omega ^2 \left(4 \omega ^2-1\right) \log ^2({\lambda_1} {\tau})-1}
$$
$$
\mbox{}
-\frac{{a} {\lambda_1} \omega  {\vartheta_2}}{{\tau}^2 \left(4 \omega ^2 \left(4 \omega ^2-1\right) \log ^2({\lambda_1} {\tau})-1\right)}
+\frac{2 (1-4\omega^2) \omega  (\lambda_3\rho_1-\lambda_1\rho_3) \log ({\lambda_1} {\tau})}{{\lambda_1} {\tau}^2 \left(4 \omega ^2 \left(4 \omega ^2-1\right) \log ^2({\lambda_1} {\tau})-1\right)}
$$
\begin{equation} 
\mbox{}
+\frac{{a} {\lambda_1} (\lambda_2\rho_1-\lambda_1\rho_2) ({\lambda_1} {\tau})^{-2 (\omega +1)} \left(4 \omega ^2 \left(4 \omega ^2-1\right) \log ^2({\lambda_1} {\tau})+1\right)}{2 \left(4 \omega ^2 \left(4 \omega ^2-1\right) \log ^2({\lambda_1} {\tau})-1\right)} 
, \end{equation} 
where $\omega$ and ${a}$ are the parameters of the considered cosmological wave model of spacetime determined by relations (\ref{MetricPrivIB}),
$\lambda_{i}$, $\vartheta_{i}$ and $\rho_{i}$ are parameters of geodesic lines determined from the initial conditions, $\tau$ is time variable - proper time on the base geodesic line.

\section{Special case II of integration of the Hamilton-Jacobi equation (${p}$ or ${q}$ equals $1/2$)}

When integrating the Hamilton-Jacobi equation for the \mbox{Shapovalov} type III space in the Bianchi type VI model for the metric (\ref{Metric}), a special case arises when ${p}$ or ${q}$ equals $1/2$ (${p}\ne {q}$). Let us consider this case and, for definiteness, without loss of generality, set

\begin{equation}
{p}=\frac{1}{2}
,\qquad
{q}=\frac{1}{2}-\omega
,\qquad
\omega\ne 0.
\end{equation}

Then from the Einstein equations we obtain a solution with one independent parameter $\omega$:

\begin{equation}
\label{metricII}
 {ds}^2=dx^0dx^1
 -\frac{1-\omega^2}{\omega^2}\left(
 {x^0}{(dx^2)}^2
 +2{a}{\left(x^0\right)}^{1-\omega}{dx^2}{dx^3}
 + {\left(x^0\right)}^{1-2\omega}{(dx^3)}^2
 \right)
,
\end{equation}
\begin{equation}
\label{alphaN2}
{a}^2=
\frac{1-2 {\omega}^2}{1-{\omega}^2}
,\qquad
0<{\omega}^2\le\frac{1}{2}
,\qquad
\det g_{{\alpha}{\beta}}=-\frac{(1-\omega^2)}{\omega^2}
\,{\left(x^0\right)}^{2(1- {\omega})}
.\end{equation}

Integration of the Hamilton-Jacobi equation for test particles of the case under consideration gives the following form of trajectories

\begin{equation} 
\label{IIX0}
x^0 ({\tau}) = {\lambda_1} {\tau} 
,\end{equation} 
\begin{equation} 
\label{IIX1}
x^1 ({\tau}) = \frac{{\lambda_3} ({\lambda_1} {\tau})^{{\omega}} \left({\lambda_3} ({\lambda_1} {\tau})^{{\omega}}-4 {a} {\lambda_2}\right)+2 {\lambda_2}^2 {\omega} \log ({\lambda_1} {\tau})+2 {\lambda_1} {\omega} {\tau}}{4 {\lambda_1}^2 {\omega}} 
,\end{equation} 
\begin{equation} 
\label{IIX2}
x^2 ({\tau}) = \frac{{a} {\lambda_3} ({\lambda_1} {\tau})^{{\omega}}-{\lambda_2} {\omega} \log ({\lambda_1} {\tau})}{{\lambda_1} {\omega}} 
,\end{equation} 
\begin{equation} 
\label{IIX3}
x^3 ({\tau}) = -\frac{({\lambda_1} {\tau})^{{\omega}} \left({\lambda_3} ({\lambda_1} {\tau})^{{\omega}}-2 {a} {\lambda_2}\right)}{2 {\lambda_1} {\omega}} 
,\end{equation} 
where $\tau$ is the proper time of the particle, $\lambda_1$, $\lambda_2$ and $\lambda_3$ are the integrals of the particle's motion determined by the initial conditions, $\omega$ and ${a}$ are the parameters of the model that meet the conditions ( \ref{alphaN2}).

The solution of the geodesic deviation equation $\eta^{\gamma}(\tau)$ for a gravitational wave (\ref{metricII}) in a privileged coordinate system is obtained from the equations (\ref{Deviation1})-(\ref{Deviation2}) in the following form
\begin{equation} 
 \eta^0 ({\tau}) = {\rho_1} {\tau}
 -{\lambda_1} {\Omega} 
  ,\qquad 
 \Omega=\vartheta_1 \lambda_1+\vartheta_2 \lambda_2+\vartheta_3 \lambda_3
   ,\qquad 
   x^0=\lambda_1\tau
, \end{equation} 
$$
 \eta^1 ({\tau}) = \frac{{\rho_1} \left(-2 {a} {\lambda_2} {\lambda_3} {\left(x^0\right)}^{{\omega}}+{\lambda_2}^2+{\lambda_3}^2 {\left(x^0\right)}^{2 {\omega}}-{x^0}\right)}{2 {\lambda_1}^3}
 $$
 $$
 +\frac{(\lambda_2\rho_1-\lambda_1\rho_2) \left({a} {\lambda_3} {\left(x^0\right)}^{{\omega}}-{\lambda_2} {\omega} \log \left(x^0\right)\right)}{{\lambda_1}^3 {\omega}}
$$
$$
\mbox{}
 -\frac{(\lambda_3\rho_1-\lambda_1\rho_3) {\left(x^0\right)}^{{\omega}} 
 \left({\lambda_3} {\left(x^0\right)}^{{\omega}}-2 {a} 
{\lambda_2}\right)}{2 {\lambda_1}^3 {\omega}}
$$
\begin{equation} 
-\frac{{\Omega} \left(-2 {a} {\lambda_2} {\lambda_3} {\left(x^0\right)}^{{\omega}}+{\lambda_2}^2+{\lambda_3}^2 {\left(x^0\right)}^{2 {\omega}}+{x^0}\right)}{2 {\lambda_1} {x^0}}+{\vartheta_1} 
, \end{equation} 
$$
 \eta^2 ({\tau})= \frac{\left({\lambda_1}^2 {\Omega}-{\rho_1} {x^0}\right) \left({\lambda_2}-{a} {\lambda_3} {\left(x^0\right)}^{{\omega}}\right)}{{\lambda_1}^2 {x^0}}-\frac{{a} (\lambda_3\rho_1-\lambda_1\rho_3) {\left(x^0\right)}^{{\omega}}}{{\lambda_1}^2 {\omega}}
 $$
 \begin{equation} 
+\frac{(\lambda_2\rho_1-\lambda_1\rho_2) \log \left(x^0\right)}{{\lambda_1}^2}+{\vartheta_2} 
, \end{equation} 
$$
 \eta^3  ({\tau})= \frac{{\left(x^0\right)}^{{\omega}-1} \left({\lambda_1}^2 {\Omega}-{\rho_1} {x^0}\right) \left({\lambda_3} {\left(x^0\right)}^{{\omega}}-{a} {\lambda_2}\right)}{{\lambda_1}^2}-\frac{{a} (\lambda_2\rho_1-\lambda_1\rho_2) {\left(x^0\right)}^{{\omega}}}{{\lambda_1}^2 {\omega}}
 $$
 \begin{equation} 
 +\frac{(\lambda_3\rho_1-\lambda_1\rho_3) {\left(x^0\right)}^{2 {\omega}}}{2 {\lambda_1}^2 {\omega}}+{\vartheta_3} 
. \end{equation} 
Here $\tau$ is the proper time of the test particle on the base geodesic line, $\lambda_{k}$, $\rho_{k}$ and $\vartheta_{k}$ (${k}=1...3$) are independent parameters, determined by the initial conditions on the geodesics, the parameters $\omega$ and ${a}$ are the parameters of the gravitational model determined by relations (\ref{alphaN2}).
  
Tidal acceleration in a gravitational wave (\ref{metricII})
takes the following form in the privileged coordinate system

$$
A^{\alpha}(\tau)=\frac{D^2 \eta^{\alpha}}{{d\tau}^2}
,\qquad
x^0 = {\lambda_1} {\tau} 
,\qquad
{a}^2=
\frac{1-2 {\omega}^2}{1-{\omega}^2}
, $$
\begin{equation} 
A^0 = 0
,\end{equation} 
$$
A^1(\tau) = 
{\vartheta_3} \left(\frac{{\lambda_1} {\lambda_3} \left({\omega}^2-1\right)}{2 {\left(x^0\right)}^2}-{a} {\lambda_1} {\lambda_2} \left({\omega}^2-1\right) {\left(x^0\right)}^{-{\omega}-2}\right)
$$
$$
+(\lambda_2\rho_1-\lambda_1\rho_2) 
\left(\frac{{\lambda_2} \left[{\omega} (1-{\omega}^2) \log \left(x^0\right)+2(2 {\omega}^2-1)\right]}{2 {\lambda_1} {\omega} {\left(x^0\right)}^2}-\frac{{a} {\lambda_3} \left({\omega}^2-1\right) {\left(x^0\right)}^{{\omega}-2}}{2 {\lambda_1} {\omega}}\right)
$$
\begin{equation} 
-\frac{{\lambda_1} {\lambda_2} \left({\omega}^2-1\right) {\vartheta_2}}{2 {\left(x^0\right)}^2}+\frac{{\lambda_3} \left({\omega}^2-1\right) (\lambda_3\rho_1-\lambda_1\rho_3) {\left(x^0\right)}^{2 {\omega}-2}}{4 {\lambda_1} {\omega}}
,\end{equation} 
$$
A^2(\tau) =   {a} {\lambda_1}^2 \left({\omega}^2-1\right) {\vartheta_3} {\left(x^0\right)}^{-{\omega}-2}
+\frac{{\lambda_1}^2 \left({\omega}^2-1\right) {\vartheta_2}}{2 {\left(x^0\right)}^2}
$$
\begin{equation} 
+\frac{(\lambda_2\rho_1-\lambda_1\rho_2) \left[\left({\omega}^2-1\right) {\omega} \log \left(x^0\right)-4 {\omega}^2+2\right]}{2 {\omega} {\left(x^0\right)}^2}
,\end{equation} 
$$
A^3(\tau) =    \frac{{a} \left({\omega}^2-1\right) (\lambda_2\rho_1-\lambda_1\rho_2) {\left(x^0\right)}^{{\omega}-2}}{2 {\omega}}
-\frac{{\lambda_1}^2 \left({\omega}^2-1\right) {\vartheta_3}}{2 {\left(x^0\right)}^2}
$$
\begin{equation} 
-\frac{\left({\omega}^2-1\right) (\lambda_3\rho_1-\lambda_1\rho_3) {\left(x^0\right)}^{2 {\omega}-2}}{4 {\omega}}
,\end{equation} 
where $\tau$ is the proper time on the base geodesic line, the parameters $\omega$ and ${a}$ are the parameters of the gravitational model determined by relations (\ref{alphaN2}), the parameters $\lambda_{k}$, $\rho_{k}$ and $\vartheta_{k}$ (${k}=1,2,3$) are independent parameters,  determined by the initial conditions on the geodesics.

The transition to the synchronous coordinate system ${\tilde x}{}^{\alpha}$, where the test particle is at rest on the base geodesic (laboratory system of a freely falling observer) can be carried out using the relations (\ref{IIX0})-(\ref{IIX3}) according to the formal rule:
\begin{equation}
x^{\alpha} \to {\tilde x}{}^{\alpha}=\left(\tau,\lambda_1,\lambda_2,\lambda_3 \right).
\end{equation}
The metric (\ref{metricII}) in the synchronous coordinate system $\tilde x{}^{\alpha}$ takes the form:
$$
\tilde g{}^{00} = 1 
,\qquad
\tilde g{}^{01} =
\tilde g{}^{02} = 
\tilde g{}^{03} = 0
, $$
$$
\tilde g{}^{11} = -\frac{{\tilde x{}^1}^2}{{\tau}^2} 
,\qquad
\tilde g{}^{12} = -\frac{{\tilde x{}^1} {\tilde x{}^2}}{{\tau}^2} 
,\qquad
\tilde g{}^{13} = -\frac{{\tilde x{}^1} {\tilde x{}^3}}{{\tau}^2} 
, $$
$$
\tilde g{}^{22} = -\frac{{\tilde x{}^1} {\omega}^2 \left({\omega}^2-1\right)^2}{{\tau} \left(\left({\omega}^2-1\right) {\omega} \log ({\tilde x{}^1} {\tau})-4 {\omega}^2+2\right)^2}-\frac{{\tilde x{}^2}^2}{{\tau}^2} 
, $$
$$
\tilde g{}^{23} = -\frac{2 {a} {\tilde x{}^1} {\omega}^2 \left({\omega}^2-1\right) ({\tilde x{}^1} {\tau})^{-{\omega}} \left(\left({\omega}^2-1\right) {\omega} \log ({\tilde x{}^1} {\tau})-3 {\omega}^2+1\right)}{{\tau} \left(\left({\omega}^2-1\right) {\omega} \log ({\tilde x{}^1} {\tau})-4 {\omega}^2+2\right)^2}-\frac{{\tilde x{}^2} {\tilde x{}^3}}{{\tau}^2} 
, $$
$$
\tilde g{}^{33} = -\frac{4 {\tilde x{}^1} {\omega}^3 \left({\omega}^2-1\right) ({\tilde x{}^1} {\tau})^{-2 {\omega}}
\log ({\tilde x{}^1} {\tau})
 \left[
{\omega} \left({\omega}^2-1\right) \log ({\tilde x{}^1} {\tau})
  -4 {\omega}^2+2 
 \right]
 }{{\tau} \left({\omega}\left({\omega}^2-1\right)  \log ({\tilde x{}^1} {\tau})-4 {\omega}^2+2\right)^2}
 $$
 $$
 \mbox{}
  -\frac{4 {\tilde x{}^1} {\omega}^2 \left({\omega}^2-1\right)
 \left(
 2 {\omega}^2-1\right)
  ({\tilde x{}^1} {\tau})^{-2 {\omega}}
 }{{\tau} \left[{\omega}\left({\omega}^2-1\right)  \log ({\tilde x{}^1} {\tau})-4 {\omega}^2+2\right]^2}
 -\frac{{\tilde x{}^3}^2}{{\tau}^2} 
. $$
The components of the geodesic deviation vector $\tilde\eta{}^{\alpha}(\tau)$ in the gravitational wave (\ref{metricII}) in the synchronous coordinate system $\tilde x{}^{\alpha}$ will take the following form

\begin{equation} 
\tilde \eta{}^0= 0 
, \end{equation} 
\begin{equation} 
\tilde \eta{}^1(\tau)= {\rho_1}-\frac{{\lambda_1} {\Omega}}{{\tau}} 
 ,\qquad 
 \Omega=\vartheta_1 \lambda_1+\vartheta_2 \lambda_2+\vartheta_3 \lambda_3
, \end{equation} 
$$
\tilde \eta{}^2(\tau)= -\frac{2 {a} {\lambda_1} {\omega} \left({\omega}^2-1\right) {\vartheta_3} ({\lambda_1} {\tau})^{-{\omega}}}{\left({\omega}^2-1\right) {\omega} \log ({\lambda_1} {\tau})-4 {\omega}^2+2}
$$
\begin{equation} 
-\frac{{\lambda_1} {\omega} \left({\omega}^2-1\right) {\vartheta_2}}{\left({\omega}^2-1\right) {\omega} \log ({\lambda_1} {\tau})-4 {\omega}^2+2}
-\frac{{\lambda_2} {\Omega}}{{\tau}}+{\rho_2} 
, \end{equation} 
$$
\tilde \eta{}^3(\tau)= -\frac{2 {a} {\lambda_1} {\omega} \left({\omega}^2-1\right) {\vartheta_2} ({\lambda_1} {\tau})^{-{\omega}}}{\left({\omega}^2-1\right) {\omega} \log ({\lambda_1} {\tau})-4 {\omega}^2+2}
$$
\begin{equation} 
-\frac{2 {\lambda_1} {\omega}^2 \left({\omega}^2-1\right) {\vartheta_3} ({\lambda_1} {\tau})^{-2 {\omega}} \log ({\lambda_1} {\tau})}{\left({\omega}^2-1\right) {\omega} \log ({\lambda_1} {\tau})-4 {\omega}^2+2}
-\frac{{\lambda_3} {\Omega}}{{\tau}}+{\rho_3} 
.\end{equation} 
Tidal acceleration in a gravitational wave (\ref{metricII}) in the synchronous coordinate system $\tilde x{}^{\gamma}$ takes the following form

$$
0<{\omega}^2\le\frac{1}{2}
,\qquad
{a}^2=
\frac{1-2 {\omega}^2}{1-{\omega}^2}<1
,$$
\begin{equation} 
\frac{D^2\tilde\eta{}^0}{{d\tau}^2}=0
,\qquad
\frac{D^2\tilde\eta{}^1}{{d\tau}^2}=0
, \end{equation} 
$$
\frac{D^2\tilde\eta{}^2}{{d\tau}^2}=
\frac{{a} {\lambda_1} \left({\omega}^2-1\right)^2 (\lambda_3\rho_1-\lambda_1\rho_3) ({\lambda_1} {\tau})^{{\omega}-2}
}{2 \left[ \left({\omega}^2-1\right) {\omega} \log ({\lambda_1} {\tau})-4 {\omega}^2+2\right]}
 $$ 
 $$ 
 \mbox{}
-
\frac{{\omega} \left({\omega}^2-1\right)^2 (\lambda_2\rho_1-\lambda_1\rho_2) \log ({\lambda_1} {\tau})
}{2 {\lambda_1} {\tau}^2 \left[\left({\omega}^2-1\right) {\omega} \log ({\lambda_1} {\tau})-4 {\omega}^2+2\right]}
 $$ 
 \begin{equation} 
\mbox{}
-\frac{{\lambda_1} {\omega} \left({\omega}^2-1\right)^2 {\vartheta_2}
}{2 {\tau}^2 \left[ \left({\omega}^2-1\right) {\omega} \log ({\lambda_1} {\tau})-4 {\omega}^2+2\right]} 
, \end{equation} 
$$
\frac{D^2\tilde\eta{}^3}{{d\tau}^2}=
{\lambda_1}^3 {\omega} \left({\omega}^2-1\right) {\vartheta_3} ({\lambda_1} {\tau})^{-2 ({\omega}+1)} 
 -\frac{{a} {\lambda_1} {\omega} \left({\omega}^2-1\right)^2 {\vartheta_2} ({\lambda_1} {\tau})^{-{\omega}}
 }{{\tau}^2 \left[\left({\omega}^2-1\right) {\omega} \log ({\lambda_1} {\tau})-4 {\omega}^2+2\right]}
 $$
 $$
 \mbox{}
 -\frac{2 {a} {\lambda_1} \left({\omega}^2-1\right) (\lambda_2\rho_1-\lambda_1\rho_2) ({\lambda_1} {\tau})^{-{\omega}-2} 
 \left[ {\omega} \left({\omega}^2-1\right) \log ({\lambda_1} {\tau})-2 {\omega}^2+1\right]
 }{\left({\omega}^2-1\right) {\omega} \log ({\lambda_1} {\tau})-4 {\omega}^2+2}
 $$
 \begin{equation} 
 \mbox{}
+\frac{{\omega} \left({\omega}^2-1\right)^2 (\lambda_3\rho_1-\lambda_1\rho_3) \log ({\lambda_1} {\tau})}{2 {\lambda_1} {\tau}^2 \left[\left({\omega}^2-1\right) {\omega} \log ({\lambda_1} {\tau})-4 {\omega}^2+2\right]} 
, \end{equation} 
where $\tau$ is the proper time on the base geodesic line, the parameter $\omega$ is the parameter of the gravitational wave determined by relations (\ref{alphaN2}), the parameters $\lambda_{k}$, $\rho_{k}$ and $\vartheta_{k}$ (${k}=1,2,3$) are 
independent parameters determined by the initial or boundary conditions for the coordinates and momenta of test particles on neighboring geodesics.

\section{Conclusion}

With this work, the authors have completed the classification of the exact solutions of the geodesic deviation equations and the test particle equations for Shapovalov type III wave spacetimes
in the Bianchi universes, initiated by research for Bianchi type IV universes \cite{Osetrin2022EPJP856} and  Bianchi type VII universes \cite{https://doi.org/10.48550/arxiv.2206.15234}.

The deviation of spacetime geodesics is the basic manifestation of gravity, which underlies theoretical models and experimental observations to detect the physical effects of gravity from the global level of the accelerated expansion of the universe to the influence of gravity on the local properties of the microwave electromagnetic background.

In this work exact models of primordial gravitational waves for Bianchi type VI universes with an analytical calculation of the deviation vector of geodesic and tidal accelerations are obtained. Usually such calculations are made by approximate methods or numerically. Exact models form the basis for the study of primordial gravitational waves. The approach of the authors demonstrated in the paper is valuable in itself as a mathematical tool and opens up additional possibilities for studying primordial gravitational waves and calculating their influence on objects and fields of the early universe.

An explicit form of coordinate transformations and a form of the metric of a primordial gravitational wave in a synchronous laboratory coordinate system are obtained, which is significant both for experimental problems and for theoretical calculations. Synchronous coordinate systems allow time synchronization at different points of space-time, which is important for measurements and comparison of measurement results at different points, incl. for accurate laser measurements, satellite measurements, etc. Calculations in the synchronous coordinate system of a freely falling observer are used for various applied problems, including satellite detection of gravitational waves, detection of the effects of their influence on the microwave radiation background  and other objects and fields of the universe.

The resulting exact models can describe the primordial gravitational waves of the universe, and also serve to debug approximate and numerical methods for detecting and analyzing the characteristics of gravitational waves by their effect on test particles.

\section*{Acknowledgments}

The authors thank the administration of the Tomsk State Pedagogical University for the technical support of the scientific project. 

The study was supported by the Russian Science Foundation, grant \mbox{No. 22-21-00265},
\url{https://rscf.ru/project/22-21-00265/}

\section*{Statements and Declarations}

\subsection*{Data availability statement}

All necessary data and references to external sources are contained in the text of the manuscript. 
All information sources used in the work are publicly available and refer to open publications in scientific journals and textbooks.

\subsection*{Compliance with Ethical Standards}

The authors declare no conflict of interest.

\subsection*{Competing Interests and Funding}
The study was supported by the Russian Science Foundation, 
\\grant \mbox{No. 22-21-00265}, 
\url{https://rscf.ru/project/22-21-00265/}





\end{document}